\documentclass[twocolumn,secnumarabic,amssymb, nobibnotes, aps, prc]{revtex4-2}

\usepackage{graphicx}
\usepackage{multirow,tabularx}
\usepackage{comment}
\usepackage{footnote}
\usepackage{amssymb,amsmath}
\usepackage{relsize}
\begin{document}
	
	\title{Fission mode identification in the $^{180}$Hg region: derivative analysis approach}

	
	\author{D. T. Kattikat Melcom}%
	\email{deby.kattikat@ganil.fr}
	\address{Present address : GANIL-SPIRAL2, Boulevard Henri Becquerel, BP 55027, F-14076, CAEN cedex 05, France}
	\affiliation{University of Bordeaux, CNRS, LP2I Bordeaux, UMR 5797, Gradignan, F-33170, France}
	
	\author{I. Tsekhanovich}%
	\affiliation{University of Bordeaux, CNRS, LP2I Bordeaux, UMR 5797, Gradignan, F-33170, France}
	
	\author{F. Guezet}%
	\affiliation{University of Bordeaux, CNRS, LP2I Bordeaux, UMR 5797, Gradignan, F-33170, France}
	
	\author{A. Andreyev}%
	\affiliation{Department of Physics, University of York, York YO10 5DD, UK}
	
	\author{K. Nishio}%
	\affiliation{Advanced Science Research Center, JAEA, Tokai, Ibaraki 319-1195, Japan}
	
	\begin{abstract}
		\begin{description}
			\item[Background] Experimental setups commonly used to study fission properties of nuclei in the exotic neutron-deficient $^{180}$Hg region are based on the time-of-flight technique for the fission-product identification. In best cases, the obtained fragments mass (FFMD) and total kinetic energy (TKE) resolution does not exceed 2 u and 6 MeV, respectively. In addition, the nuclei of interest are created via fusion reactions and at excitation energies of several tens of MeV. The deduced final FFMDs are in general structureless, which makes the identification of fission modes, along with their properties, ambiguous and author-dependent.  
			
			\item[Purpose] To develop a robust analysis approach aiming at identification of a number of major fission modes in FFMDs from fusion-fission reactions.
			
			\item[Method] FFMD data with known composition are used to calculate the $2^{nd}$ derivative which is then inspected for presence of minima.  Number of detected minima is relevant to the number of fission modes present in the FFMD and thus determines the structure of the fit function to use for the FFMD and TKE data description.  
			 
			\item[Results] The resolution effect is found to broaden the simulated FFMD and TKE data, as well as to smear structures in the 2$^{nd}$ derivatives thus limiting the sensitivity of the method to a few per-cent level in the fission-mode weight. Direct fit of the resolution-affected FFMD data is shown unable to find the asymmetric-mode positions nor to reproduce the modes weights, especially if the used fit-function composition does not correspond to the input data. The fission-modes characteristic TKE values are also not reproduced in a non-constrained fit. \textit{In contrario}, the simulated data could be safely reproduced with fit functions constructed with help of the 2$^{nd}$  derivative curves, the detected minima being the signature of number of fission modes detectable in the resolution-affected data. The power of the derivative method is demonstrated on example of real experimental FFMDs for the $^{180}$Hg nucleus. 
			
			\item[Conclusions] The derivative analysis applied to limited-resolution FFMDs appears to provide consistent results on the number and parameters of fission modes, even in cases of strong symmetric-mode dominance, i.e. for Gaussian-like FFMD shapes. The method is shown to work also on data sets with limited statistics (real experimental data with integral of a few tens of thousands of events). 
		\end{description}
	\end{abstract}
	\maketitle
	
	\section{Introduction}
	
	The process of nuclear fission, i.e. the splitting of a single nucleus into several fragments, remains one of the least comprehended fundamental phenomena in nuclear physics. Although the macroscopic Liquid Drop Model (LDM) could quantitatively explain the phenomenon of fission straight after its discovery in 1938 \cite{meitner1939disintegration}, its full understanding is yet to be achieved. The historical observation of unequal mass split in $^{236}$U, known today as the asymmetric-fission mode ($A$ mode), has been confirmed in various experiments with different actinide nuclei \cite{meitner1939disintegration}. This $A$ mode produces a double-humped fission-fragment mass distribution (FFMD) and is understood to be driven by the nucleon pairing and shell effects. The latest theoretical works refer to the octupole deformation of fragments with atomic numbers \ensuremath {Z=52} and 56 \cite{scamps2018impact, scamps2019effect} as responsible for the major mass-asymmetric splits in fission of actinide nuclei. In contrast, the symmetric-fission mode ($S$ mode) produces fragments with comparable masses and thus the relevant FFMD exhibits a Gaussian-like shape. The other global difference between the $A$ and $S$ modes is the characteristic values of the total kinetic energy (TKE) of fission products known to be lower for the $S$ mode than for the $A$-mode. This difference is understood as the consequence of more elongated fission configuration at scission point. 
	
	Actinide nuclei are known to exhibit both modes, with their relative contributions to the FFMDs governed by the excitation energy ($E^*$)  \cite{KONECKY1973329}. The dominance of the $S$ mode in the FFMDs from fission of nuclei with $A$$<$226 established experimentally in the sub-lead region, cf. \cite{aschmidt2001fission}, gave the reason to interpolate the same $S$-mode dominance also to the sub-lead region. 
	However, the double-humped FFMD obtained in a $\beta$-delayed fission ($\beta$DF) of very neutron-deficient $^{180}$Hg ($E^*$$<$10 MeV) had strongly challenged this expectation \cite{new1}. From a series of subsequently conducted experiments and theory works,  
	it is now firmly established that the mass-asymmetric fission is not a property unique to $^{180}$Hg but is relevant to a wide range of nuclei around it, cf. Refs. \cite{nishio2015excitation, kozulin2022fission,prasad2020systematics, kumar2020study, kumar2021investigation,  bogachev2021asymmetric, TSEKHANOVICH2019583, golda2019fusion, tripathi2015fission, nag2021fission, gupta2020competing, SWINTONBLAND2023137655, Prasad91.064605,Kin.85.014611, Kumar_PhysRevC.107.034614, PhysRevC.110.034613, PhysRevC.106.014616,morfouace2025asymmetric}.  
	
	Unlike the actinide region, nuclei in the vicinity of $^{180}$Hg are short-lived and can be produced in quantities sufficient for fission studies nearly exclusively by fusion technique. Here, the low mass asymmetry between fission fragments ($\sim$20 u, contrary to $\sim$40 u in actinides, cf. Ref. \cite{new1}), along with a non-perfect fission-fragment mass identification and high $E^*$ of fissioning nuclei inherent in fusion-fission reactions, introduce serious difficulties in the FFMD content determination. 
	
	In the sub-lead region, the identification of fission modes, along with the determination of their relative weight in the measured data, is typically reliant on the fit function employed to reproduce experimental FFMD and TKE distributions. This makes the obtained results highly author dependent: absence of any known prominent structures in the relevant data makes the choice of the used fit function debatable. For instance, in Refs. \cite{nishio2015excitation, nag2021fission, prasad2020systematics}, the FFMDs were described with a sum of 2-Gaussian (2G) functions centred at complementary fragment masses, and the symmetric-fission mode was not considered. Two fission-mode analysis (3G fit) of FFMDs was used, i.e., for $^{178}$Pt \cite{TSEKHANOVICH2019583, kumar2021investigation} and Bi isotopes \cite{swinton2020mass}, whereas a 5G and/or 7G fit was implemented for analysis of FFMDs from fission of $^{180,182,183}$Hg, $^{178}$Pt, and $^{182,192}$Pb \cite{bogachev2021asymmetric, kozulin2022fission}. The 5G description of the data was recently extended to two dimensions, which allowed for simultaneous fitting of the mass-TKE matrix \cite{SWINTONBLAND2023137655}. In the cited cases, the authors rely on the $\chi^{2}$ value to justify the composition of the used fit function. However, the quality of data description comes inline with the complexity of the used fit function, the latter having no obvious link to the underlying physics. In addition, none of the works published so far accounts for the impact of non-perfect experimental resolution on the obtained results.

	This paper addresses the outlined problem of the fission-mode identification in the FFMD and TKE data from fission of nuclei from the pre-lead region. We use a set of simulated data to show the effect of limited mass resolution on the capability of the fission-mode detection made with help of a classical functional-analysis approach based on the second derivative behaviour. The applicability limits of the used method are indicated. Its power is demonstrated with a set of fusion-fission data for the $^{180}$Hg nucleus taken from Ref. \cite{nishio2015excitation}. 
	
	\section{FFMD and TKE data analysis methodology}
	
	To evaluate the effect of non-ideal mass and energy resolution on the extracted FFMD and TKE distributions, a dataset with a predefined (\textit{a priori}) composition is required.
	Such datasets can be generated with a predefined number of fission modes and desired modal properties (position, width and amplitude).
	
	
	The simulated one- (FFMD) and two-dimensional (mass-TKE) datasets are convoluted with experimental resolution of 2 u and 4 u for the mass and 6 MeV and 8 MeV respectively for the TKE, which are the values most frequently cited in the fusion-fission papers in the region around $^{180}$Hg. 
	
	As a next step, the 1$^{st}$ and 2$^{nd}$ derivatives are calculated from the simulated FFMDs. The derivative data are then analysed with a set of Gaussian functions $f_{i}(A)$ describing individual FFMD components as indicated in the following equation : 
	
	\begin{equation}\label{eq1}
		\begin{gathered}
			FFMD=\sum_{i}^{}{f_{i}(A)}=\sum_{i}^{}{B_{i}}\exp\left[-\frac{(A-\overline{A_{i}})}{2\sigma_{A_i}^2} ^2\right]	\\
			f^{'}(A)=FFMD^{'}=\sum_{i}^{}{f_{i}(A)}\frac{\overline{A_{i}}-A}{\sigma_{A_i}^2} 	\\
			f^{''}(A)=FFMD^{''}=\sum_{i}^{}\frac{f_{i}(A)}{\sigma^2_{A_i}}\left[\left( \frac{A-\overline{A_{i}}}{\sigma_{A_i}}\right) ^2-1 \right]
		\end{gathered}
	\end{equation}

	Here, the functions $f^{'}(A)$ and $f^{''}(A)$ are composed of $i$ components relevant to each peak in the corresponding fission mode, with $B_i$,  $\sigma_{A_i}$ and $\overline{A_{i}}$ being the mode amplitude, width and peak position in the FFMD, respectively. The number of Gaussians to use is given by the number $i$ of minima identified in the 2$^{nd}$ derivative data. For example, $i$=5 would signify five components in the FFMD, one for the $S$-mode, two for the first $A$-mode ($A1$) and further two for the second $A$-mode ($A2$). 
	The term $f_{i}(A)$ describing the $S$- and $A$-mode components in the FFMD is present in all the three expressions of Eq.\ref{eq1}. This means that, once the composition of the fit function is determined, the FFMD composition can be obtained from the fit made on any data set (FFMD, 1$^{st}$ or 2$^{nd}$ derivative).

	However, it appears that the use of Eq.\ref{eq1}  with completely uncorrelated and unconstrained parameters may provide good overall data description in terms of $\chi^{2}$  but gives meaningless results for fission modes: the $S$-mode position may appear anywhere in the FFMD or non-identical integral area obtained for the same $A$-mode components. Therefore, it is recommended to fix the $S$-mode position at $\frac{{A_{CN}}}{2}$ and to demand identical widths, amplitudes as well as displacement from the  $S$-mode position for the same $A$-mode components.  
	
	The derivative approach cannot be directly applied to the correlated mass-TKE data.  These data are analysed with a set of two-dimensional Gaussian functions, with the number of components $i$ equal to the number of minima found in the $2^{nd}$-derivative analysis of the FFMD:

	\begin{equation}
		f(A,TKE)=\sum_{i}^{}B_{i}\exp\left[-\frac{(A-\overline{A_{i}})}{2\sigma_{A_i}^2} ^2-\frac{(TKE-\overline{TKE_{i}})}{2\sigma_{TKE_i}^2} ^2\right]
		\label{eq2}
	\end{equation}

	The feasibility of the proposed method for extracting fission modes is tested with two nuclei with a \textit{a priori} known FFMD composition: \textbf{i)} $^{240}$Pu, for which, in addition to the FFMD, the correlated mass-TKE data set can be obtained at various excitation energies using the GEF code \cite{SCHMIDT2016107}, and \textbf{ii)} a hypothetical light nucleus of $A$=180 u, whose FFMD is generated as a superposition of a selected number of fission modes, each assigned a predefined weight. 
	Finally, the $2^{nd}$-derivative method is applied to experimental FFMD data from the fusion-fission of $^{180}$Hg \cite{nishio2015excitation}.

	\section{Results}
	
	\textbf{Simulated data: $^{240}$Pu.} The FFMD of a nucleus in the classical actinide region can be best described by using one $S$- and three $A$-fission modes; here, they are denoted using the Brosa nomenclature \cite{PhysRevC.59.767} as $SL$ and $S$1-$S$3, correspondingly.  Each mode is characterised by its position in the FFMD relevant to its origin, by its specific characteristic average TKE ($\overline{TKE}$) relevant to the scission configuration, and by its relative weight $W$ indicating the mode importance in the considered FFMD.

	\begin{figure*}[htbp]
		\centering
	 	\includegraphics[scale=0.92]{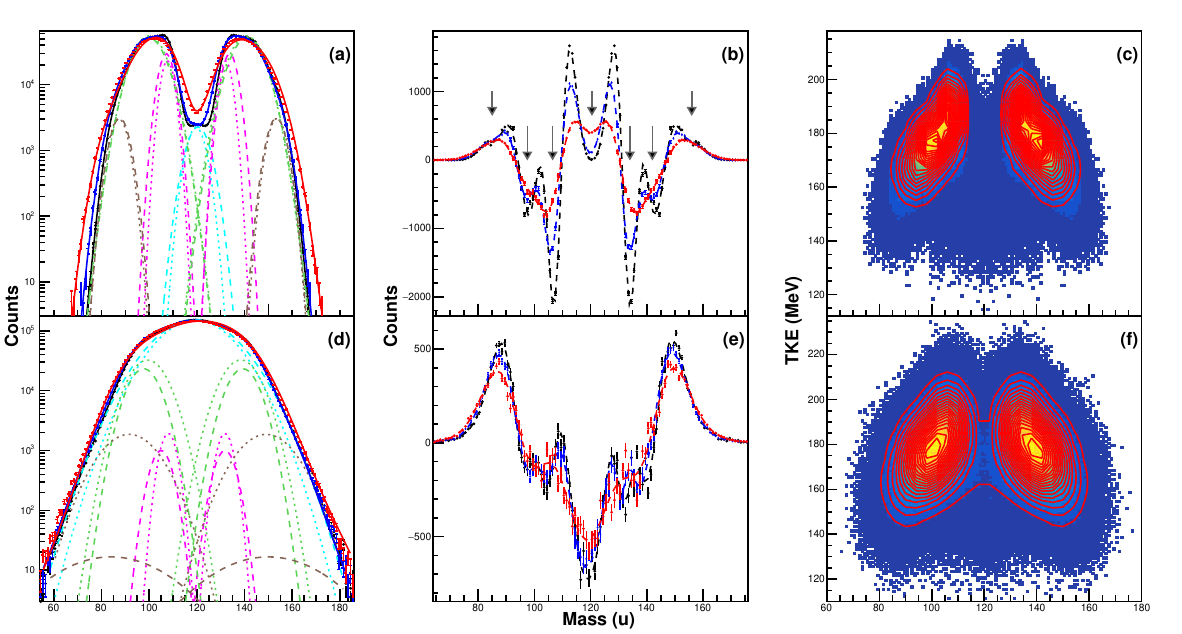}
		\caption{FFMDs (left column) and their $2^{nd}$ derivatives (middle column) for $^{240}$Pu at two different excitation energies:  $E^{*}=10$ MeV (upper row), and $E^{*}=50$ MeV (lower row). Black, blue and red lines depict, correspondingly, the simulated and the 2u- and 4u-resolution affected  data sets. The FFMD components are relevant to the fit of the simulated data set (red line) made with free (dotted lines) or constrained (dashed lines) fit function parameters. SL/S1/S2/S3 mode is shown in cyan/magenta/green/brown. The right column displays the mass-TKE matrices for $E^{*}=10$ MeV : simulated (c) and resolution-broadened data (f) (resolution 4 u for mass and 8 MeV for TKE). Here, red contour lines represent the resultant 2D fit with function from Eq.\ref{eq2}. }
		\label{Fig1}
	\end{figure*}

	Fig.~\ref{Fig1} shows how limited resolution impacts the simulated FFMD and mass-TKE data. 
	Its primary effect is the broadening of the mass and TKE distributions, which partially fills the symmetric region and reduces the peak-to-valley ratio.
	The  broadening effect is also present in the derivative data, cf. middle column of Fig.\ref{Fig1}, calculated using a sliding window on the FFMD data.  

	Presence of 7 minima is clearly seen in the simulated data set. This number determines composition of the fit function and implies presence of one $S$-mode and three $A$-modes in the considered FFMD, in accordance  with the input data. 
	
	\begin{table*}
		\caption{$^{240}$Pu : Fission-mode characteristics (weight $W$, light-fragment peak position $\bar{A_L}$ and average total kinetic energy $\overline{TKE}$) as given by GEF at different excitation energies $E^{*}$ and as found in the FFMD (1D), FFMD-TKE (2D) and the 2$^{nd}$-derivative analysis of the initial (n/ R) and resolution-corrected data (w/ R, experimental mass and energy resolution of 2 u and 6 MeV, respectively).  $\chi^2$ is the data description quality. The units are \% for $W$, u for $\bar{A_L}$ and MeV for $E^{*}$ and $\overline{TKE}$. } 
		
		\label{table_1}
		\begin{ruledtabular}
			\begin{tabular}{c c | ccc | c c c  | c c c c c c c c c c }
				
				\multirow{3}{*}{} & \multirow{3}{*}{} & \multicolumn{3}{c|}{Input data} & \multicolumn{3}{c|}{n/ R} & \multicolumn{8}{c}{w/ R}  \\
				\cline{3-18}
				\multirow{2}{*}{$E^*$} & \multirow{2}{*}{Mode} & \multicolumn{3}{c|}{} & \multicolumn{3}{c|}{1D}  & \multicolumn{3}{c}{1D}  & \multicolumn{3}{c}{Second derivative}  & \multicolumn{3}{c}{2D} \\

				\cline{6-18}
				& & $W$ & $\bar{A_L}$ & $\overline{TKE}$ & $W$ & $\bar{A_L}$ & $\chi^2$ & $W$  & $\bar{A_L}$ & $\chi^2$  & $W$ & $\bar{A_L}$ & $\chi^2$ & $W$ & $\bar{A_L}$ & $\overline{TKE}$ & $\chi^2$ \\
				
				\hline
				10  & $SL$ &   3.0 & 120.0 & 165(4) & 1.3   & 120.5 &5.74 & 1.1  & 120.0  &1.12& 1.7  & 119.6  &1.77&2.5 & 119.9 & 155.0 & 13.8\\	
				& $S$1 & 20.2 & 106.9 & 189(4) & 22.0 & 107.2 &       & 21.0& 106.7 &        &22.3& 106.6 &       &52.1& 104.8 & 184.7 & \\
				& $S$2 & 69.8 &  99.6 & 177(4) & 73.5 & 99.8   &       & 75.0 &  99.4 &        &73.0&  99.0 &       &1.8  & 110.0  & 167.0 & \\ 
				& $S$3 &   6.0 &  88.8 & 162(5) & 3.1  & 88.0   &       & 2.8   &   87.4 &        &  2.9 &   87.4 &       &43.6& 95.9 &170.0 & \\ 
				
				\hline	
				30  & $SL$ & 64.8 & 119.7  & 171(6) & 95.6 & 120.5& 1.53& 81.5& 120.0&0.98& 75(14)   & 119.7   & 1.33& 27.2 & 120.1 & 163.1 & 0.93\\	
				& $S$1 &   3.7  & 106.9 & 189(4)&   0.3 & 115.5&         &  5.4 & 116.5&       & 2.4(24) & 104.0  &        & 55.6 & 113.3 & 179.1 & \\
				& $S$2 & 30.4 &  99.2 & 178(3) &   3.7 & 101.3&         & 11.5 & 104.4&       &12.5(35) & 100.0  &        &  7.3  &  90.2 & 165.0 & \\
				& $S$3 &    1.1  &  87.9 & 163(3) &  0.4 &  90.6&         &   1.6 & 96.5 &       &  9.6(29) & 80.8   &        & 9.9  & 99.6  & 170.0 & \\
				
				\hline		
				50  & $SL$ & 85.4 & 118.8 & 173(3)  & 23.2& 119.3 & 7.18& 19.5 &118.8&5.27&83.6(23)& 118.8&1.39& 41.3 & 119.6 & 172.9 & 1.59\\	{\tiny }
				& $S$1  &   1.1   & 105.7 & 189(3) &63.8& 112.2 &         & 66.3 &112.6&       &   0.5(2) & 105.2&       & 19.4 & 110.7 & 183.7 & \\
				& $S$2  & 13.2  &  98.3 & 179(5) & 13.0&   99.1 &         &  13.8& 99.4&        &15.2(9) & 98.1 &       & 29.1 & 102.7 & 170.6 & \\
				& $S$3 &   0.3  &  87.4 & 164(3) & 0.01&    91.1 &         &  0.3  & 93.8&        &  0.8(3)& 92.0  &       & 10.3 &  99.4 & 157.4 & \\
				
				\hline		
				70   & $SL$ & 92.9 & 117.9  & 174(4) & 21.5& 118.4& 3.30 & 15.9& 117.9&2.20&94.0(33)& 118.2& 1.63&   0.1 & 118.4 & 166.6 & 1.16\\	{\tiny }
				& $S$1 &  0.6  & 104.8 & 190(3) & 74.0& 111.3&          & 79.2&109.3&        &   0.3(1) & 104.7&       & 81.5 & 110.0 & 175.8 & \\
				& $S$2 & 6.4  &  97.3  & 180(3) &   4.4 & 97.1 &          &  3.6 & 96.6&        &   4.4(5) & 97.0 &        & 10.6 &  95.6 & 173.6 & \\
				& $S$3 &  0.1  &  86.3 & 164(3) & 0.02 & 90.0&          &   1.3 & 79.3&        &  1.3(23) & 81.0 &         &   7.8 &  92.0 & 159.8 & \\
				
			\end{tabular}
		\end{ruledtabular}
	\end{table*}

	From the one- (1D) and two-dimensional (2D) fits with the 7G fit function on the simulated and resolution affected data, the 
	$S$- and $A$-mode parameters are extracted and parameters of the $S$- and $A$-modes are obtained and can subsequently be compared with the input values.  Table \ref{table_1} demonstrates this sort of comparison made for a range of $E^*$ from 10 to 70 MeV. The following can be stated:
	
	\begin{itemize}
		\item For $E^*$$\geq$30 MeV, a direct FFMD fit (1D in Table \ref{table_1}) does not correctly reproduce either the mode positions or their respective weights, the disagreement with input data being sometimes stronger for the resolution-affected data. The striking feature here is the suppression of the $SL$-mode, in favour of the $S1$-mode. 
		
		\item Globally, free fit on the resolution-affected 2$^{nd}$ derivative data correctly reproduces the dominant mode positions $\bar{A_L}$ and weights $W$. $\bar{A_L}$ of minor modes are less satisfactory reproduced, as seen on example of $S3$, for which the depth of the minima (at $A$=88/152 u, $W$=6\% at $E^*$=10 MeV) is affected by the decreasing resolution, cf. Fig.~\ref{Fig1}(b). With increasing $E^*$, the $S3$-mode weight drops below 1\% which makes it be hardly detectable in the 2$^{nd}$ derivative data even at perfect resolution. 
		
		\item The 2D free fit fails to reproduce the $\overline{TKE}$ and $W$ data even for dominant fission modes. Using the 2$^{nd}$ derivative minima to fix the $\bar{A_L}$ in the 2D fit function helps in the $\overline{TKE}$ determination (results not shown in Table \ref{table_1}) but still does not deliver correct $W$ values.   
	\end{itemize}

	It is worth noting that the simulated data exhibit high statistical precision ($>10^5$), allowing for determination of $W$ with uncertainty below 1\%. We also stress that both the 1D and 2D fits are sensitive to the specified fit boundaries, as also emphasized in Ref. \cite{SWINTONBLAND2023137655}. The associated uncertainty can be gauged by varying the boundary conditions, revealing the value surpassing the statistical one by a factor of 4. Consequently, an overall uncertainty of approximately 4\% can be assigned to the 1D and 2D $W$ values in Table \ref{table_1}.

	\textbf{Simulated data: A=180}. The FFMD of this hypothetical nucleus can be simulated by assuming different $S$ and  $A$ mode combinations. In the simulation, we focus on the two selected combinations: one $S$ plus one $A$ mode and one $S$ plus two $A$ modes, the $A$ modes displaced by 10 and 24 u with respect to the centroid $\frac{A_{CN}}{2}=90$.

	\begin{figure}[htb]
		\centering
				\includegraphics[trim=20 0 30 0,scale=.22]{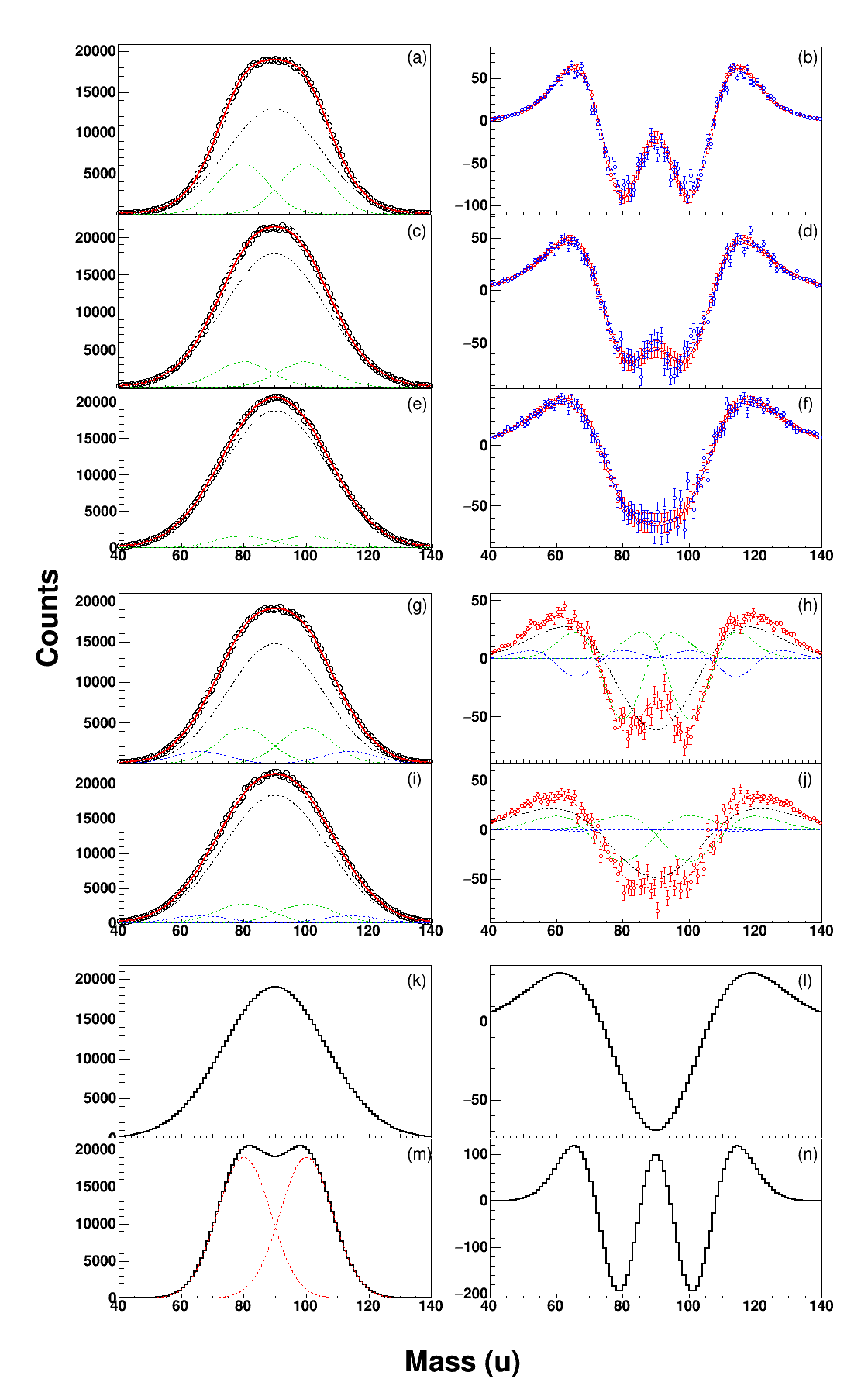}
		\caption{ \textbf{Four first rows}: simulated FFMDs (black symbols, left column) affected by resolution of 2 u and their 2$^{nd}$  derivatives (right column) for different combinations of $S$ and $A$ modes, refer to Table \ref{table_2} for input data. Panels (a,c,e): sum results of the 3G function FFMD fit (red line), panels (a—f): fission modes components shown with black and green broken lines for the $S$ and $A$ modes, respectively.  Resolution effect is demonstrated with the 2$^{nd}$ derivative data: blue/red symbols stand, correspondingly, for the initial and resolution-affected (where a random function was used to simulate the resolution impact) data sets.  Broken black, green and blue lines in (g-j) are respectively the $S$, $A1$ and $A2$ mode components from the fixed parameter fit. \textbf{Two bottom rows}: pure $S$ and pure $A$ mode FFMDs and their 2$^{nd}$ derivatives. 
		}
		\label{Fig2}
	\end{figure}

	\begin{table*}
		\caption{FFMD and derivative analysis for $A=180$:  FFMDs simulated with different contributions from 2 or 3 fission modes and convoluted with 2 u of resolution. $S$, $A1$ and $A2$ are the symmetric and two asymmetric modes, $W$ (in \%) denoting their weight. $\bar{A_L}$ and $\sigma$, in units of u, refer to the light-fragment position and width of fission modes.}
		\label{table_2}
		\begin{ruledtabular}
			\begin{tabular}{c|| c c c| c c c c  | c c c c}
				
				\multirow{3}{*}{Mode} & \multicolumn{3}{c|}{Input data} & \multicolumn{8}{c}{FFMD / second derivative fit results}   \\
				\cline{2-12}
				&\multicolumn{3}{c|}{} 
				&\multicolumn{4}{c|}{Fit function: 3 Gauss ($S$+$A1$)} 
				&\multicolumn{4}{c}{Fit function: 5 Gauss ($S$+$A1$+$A2$)}  \\
				&$W$ & $\bar{A_L}$  & $\sigma$\footnotemark[2]  & $W$ & $\bar{A_L}$  & $\sigma$ & $\chi^2$ & $W$ & $\bar{A_L}$ & $\sigma$ & $\chi^2$  \\
				
				\hline
				$S$  &65.2& 90 & 15 & 66 / 72  &       90.0  & 15.1 / 15.3 & 1.03 / 1.93& 68.8 / 53.9 & 90.0        & 14.9 / 17.1 & 0.90 / 1.35\\
				$A$1 &34.8& 80 & 8  & 34 / 28  & 80.0 / 80.0 & 8.2 / 8.3   &            & 30.8 / 46.0 & 79.8 / 80.8 &  8.1 / 9.2  &      \\ 
				$A$2 & —  & —  & —  &          &             &             &            & 0.4 / 0.1   & 68.3 / 59.1 & 12.7 / 4.4  &          \\ 			
				\hline 
				$S$  &82.9& 90 & 16 &  83 / 83 &        90.0 & 16.1 / 16.5 & 1.03 / 1.72& 65.3 / 62.1 & 90 .0       & 15.6 / 18.6 & 0.89 / 0.90 \\
				$A$1 &17.1& 80 & 8.5&  17 / 17 & 80.0 / 80.0 & 8.8 / 9.0   &            & 26.6 / 37.2 & 80.7 / 81.1 &  9.2 / 10.1 &           \\ 
				$A$2 & —  &  — & —  &          &             &             &            & 8.1 / 0.7   & 72.0 / 62.1 & 13.3 / 7.5  &           \\ 	 
				\hline
				$S$  &91.4& 90 &16.5&  91 / 92 &       90.0  & 16.6 / 17.1 &0.85 / 1.68 & 59.4 / 80.3 & 90.0         & 14.8 / 18.18& 0.91 / 1.39\\
				$A$1 &8.6 & 80 & 9  &  9 / 8   & 80.0 / 80.0 &  9.2 / 9.1  &            & 34.8 / 19.7 & 79.4 / 81.4  & 11.3 / 10.9 &            \\ 
				$A$2 & —  & —  & —  &          &             &             &            &    5.8 / 0.1& 61.7 / 58.5  & 11.5 / 12.0 &             \\ 
				\hline
				$S$  &70.1& 90 &15  & 25.6 / 47.8 &      90.0   & 15.5 / 16  & 1.80 / 1.5 & 69.6 / 87   &  90.0       & 15.1 / 16  & 0.77 / 1.1\\
				$A$1 &22.4& 80 &8   & 74.4 / 52.2 & 80.1 / 79.6 & 12.3 / 11.7&            & 22.7 / 10.3 & 80.0 / 80.0 & 8.3  / 8.3  &      \\ 
				$A$2 &7.5 & 66 &8   &  —          & —           & —          &            & 7.7 / 2.7   & 66 / 66     & 8.3  / 7.9    &      \\ 	                 
				\hline
				$S$  &84.0& 90 &16.5& 68 / 66 &      90.0   & 16.7 / 18.1 & 1.10 / 1.42 & 81.4 / 85.3 &  90.0      & 16.5 / 18.1 & 1.16 / 1.93\\
				$A$1 &11.4& 80 &8.5 &  32 / 34& 78.7 / 80.1 & 12.8 / 12.0 &             & 13.3 / 14.3 & 80.0 / 80.0& 9.1 / 11.5 &      \\ 
				$A$2 &4.6 & 66 &8.5 &  —      & —           & —           &             & 5.3  / 0.4  & 66 / 66    & 9.4 / 9.9  &      \\

			\end{tabular}
		\end{ruledtabular}
		\footnotetext[2]{Accounts for experimental resolution of 2 u}
	\end{table*} 

	The simulated data address the FFMD shapes commonly observed in fusion-fission reactions in the sub-lead region; the input data on fission modes are summarized in Table \ref{table_2}. The point of interest here is to make analysis of the resolution-affected FFMDs displayed in Fig.\ref{Fig2}. 

	As seen above, the resolution effect manifests as a flattening of structural features in the FFMD and its derivative data.  The right column of Fig.\ref{Fig2} demonstrates the power of the $2^{nd}$ derivative analysis: 
	it appears possible to identify presence of $A$ modes in the FFMDs strongly dominated by the $S$ mode, cf. panels (f,j). In these extreme cases, the presence of the $A1$ mode is evidenced either by a shallow minimum at the $\frac{A_{CN}}{2}$, if compared to the pure $S$-mode derivative from panel (l). The $A1$ mode position can in this case be fixed and fitted from the minima of the more structured FFMDs, such as in panels (b,d), where the position can be read directly by eye.
	Panel (j) also illustrates the sensibility of the method to detect the $A2$ contribution to the FFMD, which is kept here on the detection limit. The minimum visible at $A$$\sim$66 is expected to become flatter or be completely blurred if the data are measured with resolution worse than 2 u and when the $A2$ mode contribution is less than 4\%, similar to the $S3$ mode in Fig.\ref{Fig1}(b). 
	 
	To clarify on how the choice of the fitting function may affect the determination of the major fission-mode parameters, the 3G and 5G functions were used to fit data from Fig.\ref{Fig2}. For each fit, the $S$ mode position was kept fixed at 90 u, whereas the $A$ modes parameters were kept free but identical for each particular $A$ mode. Table \ref{table_2} compares the obtained results with the input data.

	\begin{itemize}
		\item  True fit function composition: 
		Fitting both the FFMD and the derivative datasets including resolution effects provides mutually consistent results and successfully reproduces the input parameters ($\bar{A_L}$, $\sigma$ and $W$).
	
		\item Wrong fit function composition: 
		Both the direct FFMD fit and the $2^{nd}$ derivative-data fit fail to reproduce the predefined input parameters. The resulting discrepancy between the two sets of mode parameters provides a useful indicator of an inadequate fitting function.
		

	\end{itemize}

	\textbf{Real data: $^{180}$Hg}. In this section, the derivative approach is applied to experimental data from the study in Ref. \cite{nishio2015excitation}. Notably, this work was the first, after the discovery of asymmetric fission of $^{180}$Hg at ISOLDE \cite{new1}, to employ fusion–fission reactions for fission studies in the $^{180}$Hg region. It is stated in Ref. \cite{nishio2015excitation} that the resulting FFMDs "\textit{could be well reproduced with a single asymmetric fission mode}". 
	This tacitly implies that the $S$-mode does not appear to play a significant role in the fission of $^{180}$Hg, despite the excitation energy exceeding 30 MeV.

\begin{figure}[htb]
	\centering
	\includegraphics[width=1.0\linewidth]{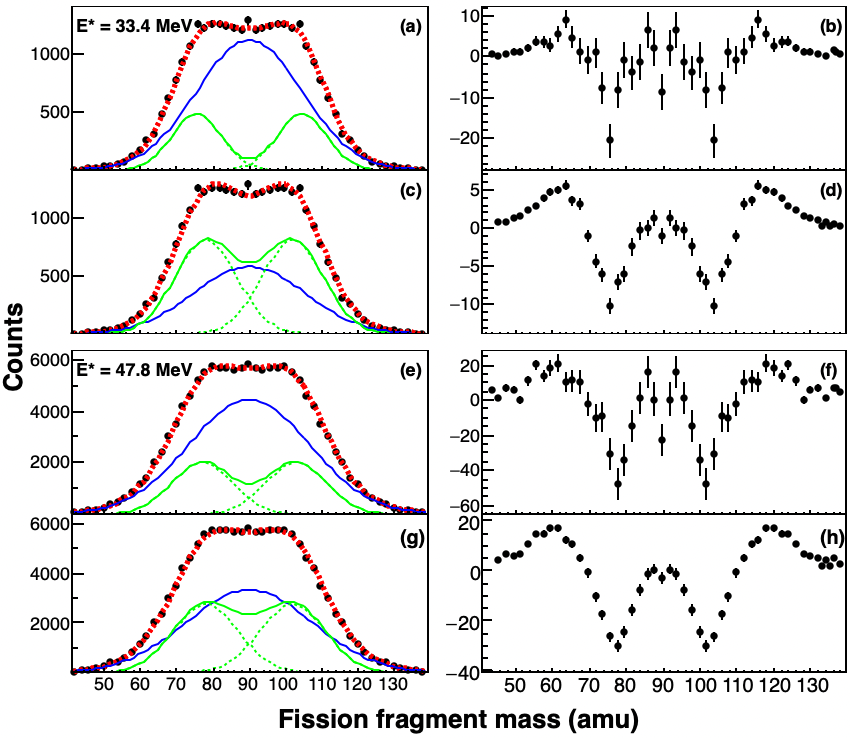}
		\caption{Experimental FFMD for $^{180}$H at two different excitation energies taken from Ref. \cite{nishio2015excitation} and their  derivatives calculated with window length of 3 u (panels (\textbf{b,f})) and 5 u (panels (\textbf{d,h})). The minima identified at $A$=78, 90 and 102 u in the 2$^{nd}$ derivative data determine the fit function composition as $S$ plus one $A$ mode.  Panels (\textbf{a,e}) and (\textbf{c,g}) display, respectively, the individual mode components from the FFMDs fit with free and fixed $A$-mode peak positions.  Blue/green lines indicate the $S$/$A$-mode components, red line is their sum.  The mass resolution was 2.4 u.}
	\label{Fig3}
\end{figure}

	Fig.\ref{Fig3} shows the $^{180}$Hg FFMD data (left column) from \cite{nishio2015excitation} analysed for minima in the 2$^{nd}$ derivative curves (right column) for the cases with best statistics ($E^*$=33.4 - 47.8 MeV). True experimental FFMDs are limited in statistics and show scattering of data around a smooth shape. This scattering affects the 2$^{nd}$ derivative curve but can be minimised with the choice of the derivative window length, cf. panels (b,d) and (f,h). Two dip minima are found present at both $E^*$, their positions do not change with $E^*$ and can be read directly from panels (b,f) or (d,h). Presence of the $S$ mode is evidenced by minimum at $A$=90, and also from comparison with the pure-mode shapes (l,n) of Fig.\ref{Fig2}. With no other minima clearly present in the  2$^{nd}$ derivative data, the fit function of Eq.\ref{eq1}  is then determined to consist of three components.
	
	The FFMDs have consequently been fitted with the 3G function. The fits were made with both free and fixed position of the $A$ mode, cf. Table \ref{table_3} and Fig.\ref{Fig3}  for the results. A juxtaposition of these results gives rise to the following observations:

	\begin{itemize} 
		\item  FFMD fit with free $A$-mode parameters shows a drift of the mode position with $E^*$, in contrast to the 2$^{nd}$ derivative data, as well as an irregular change in width $\sigma$ and weight $W$.  Conversely, the FFMD fit with fixed $A$-mode positions results in increasing $A$- and $S$-mode weights $W$ and widths $\sigma$ as a function of $E^*$, a trend that seems physically plausible.
		Moreover, such a fit does not lead to the loss of accuracy in the data description, cf. reduced $\chi^2$ values in Table \ref{table_3}.  
		
		\item The found disagreement between the two used fit approaches is expected to vanish if the FFMD event number becomes sufficiently high ($\sim10^5$).
	\end{itemize}

	\begin{table}
		\caption{Analysis of $^{180}$Hg data from Ref. \cite{nishio2015excitation}: Fission-mode weight $W$ (\%), position $\bar{A_L}$  (u) and width $\sigma$ (u) from the FFMD fit with a 3G fit function with free and constrained $A$-mode peak positions. $\chi^2$ is the fit quality. }.
		\label{table_3}
		\begin{ruledtabular}
			\begin{tabular}{c c|| c c c c | c c c c }
				
				\multirow{2}{*}{$E^{*}$}  & \multirow{2}{*}{M}  & \multicolumn{4}{c|}{Free $A_L$ position} & \multicolumn{4}{c}{Fixed $A_L$ position}    \\
				
					&         & $W$ &$\bar{A_L}$ & $\sigma$& $\chi^2$ & $W$ & $\bar{A_L}$ & $\sigma$& $\chi^2$  \\

			\hline
			33.4&$S$  & 90(8)  & 90.0     & 19(2)  & 1.59& 42(3) & 90.0& 15.4(1)& 1.64 \\
					&$A$1& 10(7) & 78.9(1) & 9.3(2)&         & 58(3) & 78.0& 8.5(1)  &       \\  	             
			\hline
			47.8&$S$  & 77(4) & 90.0     & 16.2(1)& 4.23& 51(3)  & 90.0& 17.0(1)&4.79  \\
					&$A$1& 23(2) & 75.8(4)& 7.7(3) &         & 49(3)  & 78.0 & 9.4(1)&         \\

			\end{tabular}
		\end{ruledtabular}
	\end{table}

	\section{Conclusions}
	
	The study addresses the challenge of fission-mode identification in FFMD and TKE data measured with finite experimental resolutions, typically on the order of several atomic mass units (u) in mass and several MeV in energy.
	The influence of experimental resolution has been studied as a function of excitation energy, using two nuclei with markedly different properties as representative cases:
	 a typical actinide nucleus with a well-structured FFMD ($^{240}$Pu) and a hypothetical $A$=180 nucleus with a Gaussian-like FFMD shape. Globally, the resolution effect smears the structures and broadens the FFMD and TKE data thus making fission modes with contribution on a few percent level difficult to detect. 
	
	It is shown that the FFMD fit with free parameters is unable to correctly reproduce the input data on the mode positions and weights, especially if adopted fit function does not correspond to the true FFMD composition. 
	The wrong mode position determination often leads to an overestimated contribution of the asymmetric mode, on expenses of the symmetric one. 
	In addition, fit of the FFMD-TKE matrix is found to fail in the $\overline{TKE}$ determination of fission modes, the disagreement can be as large as 10 MeV even for dominant modes. This failure can be explained by presence of multitude of local minima on the mass-TKE landscape, whereas the shift of the mode positions from the true values is likely to assign to the fit algorithm aimed at the best data description. 
	
	Furthermore, the direct FFMD fits of the $A$=180 nucleus using 3G and 5G fitting functions show that, without \textit{a priori} knowledge of the appropriate functional composition, the extracted results may lead to incorrect physical results: increase of the $A$-mode contribution with increasing excitation energy.

	 Simple counting of number of extrema in the 2$^{nd}$ derivative data is shown to be able to resolve the ambiguity in the determination of the FFMD fitting function composition. 
	 Although the function derived by this procedure can be employed directly on the 2$^{nd}$ (or 1$^{st}$) derivative datasets to retrieve the FFMD composition, the optimal parameter precision is achieved when it is fitted directly to the FFMD dataset. 
 
	 If applied to real experimental data ($^{180}$Hg from Ref. \cite{nishio2015excitation}), the 2$^{nd}$ derivative analysis shows itself capable of detecting, in a simple and undoubted way, the presence of both $S$ and $A$ modes in the FFMD data, thus resolving the question on the $S$-mode presence in the data raised by the authors. 
     The $A$-mode weight extracted with a fixed $A$-mode position exhibits the anticipated increase with excitation energy, lending additional credibility to the analysis method applied to datasets with limited statistics.

	\section{Acknowledgments}
	The authors (K.M.D.T. and K.N.) acknowledge support from French Ministry for higher Education, Research and Innovation (MESRI) and the Bordeaux University.

	\bibliographystyle{elsarticle-num}
	\bibliography{Fission-modes-identification.bib}
	
\end{document}